# Inquiries into the Nature of Free Energy and Entropy In Respect to Biochemical Thermodynamics

**Clinton D. Stoner**

*Department of Surgery, Ohio State University, Columbus, Ohio 43210, U.S.A.*

**ABSTRACT:** Free energy and entropy are examined in detail from the standpoint of classical thermodynamics. The approach is logically based on the fact that thermodynamic work is mediated by thermal energy through the tendency for nonthermal energy to convert spontaneously into thermal energy and for thermal energy to distribute spontaneously and uniformly within the accessible space. The fact that free energy is a Second-Law, expendable energy that makes it possible for thermodynamic work to be done at finite rates is emphasized. Entropy, as originally defined, is pointed out to be the capacity factor for thermal energy that is hidden with respect to temperature; it serves to evaluate the practical quality of thermal energy and to account for changes in the amounts of latent thermal energies in systems maintained at constant temperature. With entropy thus operationally defined, it is possible to see that $T\Delta S^\bullet$ of the Gibbs standard free energy relation $\Delta G^\bullet = \Delta H^\bullet - T\Delta S^\bullet$ serves to account for differences or changes in nonthermal energies that do not contribute to $\Delta G^\bullet$ and that, since $\Delta H^\bullet$ serves to account for differences or changes in total energy, complete enthalpy-entropy ($\Delta H^\bullet$-$T\Delta S^\bullet$) compensation must invariably occur in isothermal processes for which $T\Delta S^\bullet$ is finite. A major objective was to clarify the means by which free energy is transferred and conserved in sequences of biological reactions coupled by freely diffusible intermediates. In achieving this objective it was found necessary to distinguish between a 'characteristic free energy' possessed by all First-Law energies in amounts equivalent to the amounts of the energies themselves and a 'free energy of concentration' that is intrinsically mechanical and relatively elusive in that it can appear to be free of First-Law energy. The findings in this regard serve to clarify the fact that the transfer of chemical potential energy from one repository to another along sequences of biological reactions of the above sort occurs through transfer of the First-Law energy as thermal energy and transfer of the Second-Law energy as free energy of concentration.

## INTRODUCTION

Despite the assurance of the First Law of Thermodynamics that energy is invariably conserved, we often concern ourselves with the conservation of energy. This apparent inconsistency obviously must arise from the fact that the ambient thermal energy into which high-grade thermal energy and most other forms of energy tend to degrade has relatively little 'available work potential'. And it must be this so-called *free* energy rather than an actual energy that we ordinarily seek to conserve, 'actual energy' referring to any of the various energies recognized by the First Law.

The primary objectives in pursuing this study were to achieve fully operational understandings of free energy and its close relative entropy and thereby to clarify apparent inconsistencies of the above sort and to deal with a number of other problems that tend to make the laws and fundamental equations of thermodynamics generally difficult to interpret and understand. Some of the problems and their persistence appear to derive from the common practice of viewing entropy as a measure of disorder. Entropy as disorder is not measurable as such and thus is not operational. Whatever the case, the findings here suggest that several of the difficulties that commonly plague the interpretation of thermodynamic phenomena can be resolved by identifying and distinguishing between the various energies involved and taking into consideration the practical qualities of the actual energies in respect to free energy.

The original objective was simply to clarify the means by which free energy is transferred and conserved in sequences of stationary-state biological reactions coupled by freely diffusible intermediates. The reported findings in this regard serve to emphasize the need to acknowledge the existence of an intrinsically mechanical, phantom-like free energy that can disappear without a trace and thereby appear to be free of actual energy.

**Energy, Temperature, Work, Work Potential, and Free Energy.** Energy is usually defined in terms of a capacity of something to do work. Actual energy of matter may be classed broadly into two interconvertible varieties: energy of motion (kinetic energy) and energy of constrained motion (potential energy). Energy of motion can be subdivided into directed and undirected varieties. Disregarding the radiant form, thermal energy is an actual energy of undirected (random) motion of the individually mobile particulate constituents of macroscopic amounts of matter and is a mechanical kind of energy into which all other forms of actual energy tend to convert. Temperature is a property of macroscopic amounts of matter and serves to gauge the intensity of the thermal energy. Thermal energy transfer occurs spontaneously and net transfer along a gradient of temperature is a one-way process, occurring only from higher to lower temperature. In consequence, macroscopic amounts of matter in thermal contact with one another tend to be at the same temperature, a fact of sufficient fundamental importance to merit belated designation as the Zeroth Law of Thermodynamics.

Work may be defined roughly as any activity that is energetically equivalent to lifting a weight. Since it exists only at the time it is being performed, work is generally viewed both as a nonthermal actual energy in transit between one form or repository and another and as a means of nonthermal actual energy transfer. We shall be concerned here only with 'thermodynamic work' (*i.e.*, only with work in which the transit of nonthermal actual energy between forms or repositories occurs through intermediary thermal energy). Accordingly, 'conservative work', such as that done in direct conversions between gravitational potential energy and directed motional energy of matter, shall be ignored. Transfers of actual energy in thermodynamic work processes require specific mechanisms that depend on the nature of the energy. Thermal energy transfer occurs by conduction, convection, and radiation.

Consistent with the reciprocity in the definitions of work and energy, work potential shall be considered here to be the total potential of an energy for doing work. Although in principle all actual energies are equivalent to their work potentials and thus are equivalent with respect to work potential, they differ with respect to available work potential and thus differ with respect to quality, the quality being higher the greater the availability of the work potential. The availability of the work potential of an actual energy depends on the nature of the energy and on the conditions under which work is done. Whereas the work potential of thermal energy is completely unavailable at constant temperature, that of all other forms of actual energy can in principle become available at constant temperature as a result of (*i*) the tendency of nonthermal actual energy to convert spontaneously into thermal energy of quality (temperature) exceeding that of the ambient thermal energy and of (*ii*) the capacity of relatively high-quality thermal energy to do mechanical work through its tendency to distribute spontaneously and uniformly within the accessible ambient space. In the course of any work thus done, actual energy and work potential are invariably conserved, whereas energy quality, or available work potential (free energy), is invariably consumed. As will be emphasized below, the amount of free energy consumed in the course of a thermodynamic work process depends greatly on the magnitude of the differential in quality between the intermediary and ambient thermal energies and thus on how fast the work is done.

Since it is possible in principle for thermodynamic work to be done under conditions of the rate and the differential in thermal energy quality being extremely small, any nonthermal actual energy can be considered to possess in principle an amount of free energy equivalent to the amount of the energy itself. In consequence, all actual energies other than thermal energy may be viewed as being completely interconvertible through mechanical work at ordinary ambient temperatures and thus may be viewed equally as energy of high quality. However, owing to differences in barriers to conversion, to differences in complexity of required mechanisms, and to consequent unavoidable differences in losses of free energy in real conversion processes, the qualities of the various nonthermal actual energies differ from the standpoint of practical processes. Because gravitational potential energy can be converted directly and unimpededly into mechanical work, the energy conserved in the course of lifting a weight is generally viewed as the form of nonthermal actual energy having the highest quality.

Although completely unavailable at constant temperature, the work potential of thermal energy can be realized at the expense of a decrease in the temperature of the energy, and thus for thermal energy having a particular temperature the work potential may be viewed as being potentially available. However, because complete conversion of a given amount of thermal energy into a nonthermal actual energy would require a decrease in the temperature of the given amount to absolute zero, the potential availability of the work potential of thermal energy is quite limited and that of ambient thermal energy at ordinary ambient temperatures is practically nil.

**THERMODYNAMICS OF GASES**

The fundamental aspects of thermodynamics are based largely on energy changes associated with changes in the state properties pressure, volume, and temperature of an ideal gas in accordance with the ideal gas equation: $PV = nRT$. In consequence, the nature of free energy and its relationship to entropy are best seen by examining the predictions of the laws of thermodynamics in respect to compression and expansion of a gas of this kind. Such will be done here by considering an ideal gas confined within a rigid cylinder of the sort shown in Figure 1. The walls of the cylinder are assumed to be permeable to thermal energy and in contact with a thermal reservoir that is sufficiently large and conductive to validate the assumption that its temperature remains constant despite transmissions of thermal energy between it and the gas. The gas is assumed to be confined to the cylinder and separated from a vacuum by a massless and frictionless piston by which it can be compressed and allowed to expand.

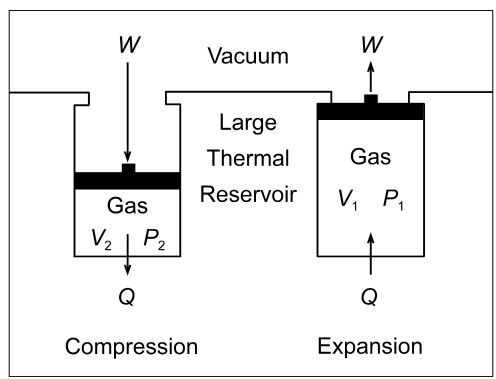

**Figure 1. Compression and expansion of an ideal gas in a rigid-cylinder, piston system.** The piston is assumed to be massless and frictionless and the temperature of the thermal reservoir is assumed to be constant. $W$ = an amount of mechanical work; $Q$ = an amount of thermal energy; $V$ = volume of the gas; $P$ = pressure of the gas.

It is important to note that, since ideal gases and solutes in ideal solution are virtually identical in respect to mechanical thermodynamic properties at normal ambient temperatures and pressures [1], the Figure 1 system is also relevant to the thermodynamics of dissolved substances. Thus, although the kinetics would be much different, the same thermodynamic relationships would apply if (*i*) the gas of Figure 1 were a solute in ideal solution, (*ii*) the vacuum were replaced by the pure solvent, and (*iii*) the piston were selectively permeable to the solvent. If such were the case, the concentration and osmotic pressure of the gas as a solute would correspond to the concentration and pressure of the gas as a gas, and the thermodynamic relationships would be largely independent of the pressure on the solvent within the normal atmospheric range of pressure.

According to the ideal gas equation and the First Law of Thermodynamics, if an amount of work $W$ were done on an ideal gas in compressing it from the maximum volume $V_1$ to a volume $V_2$, an amount of thermal energy $Q$ equivalent to the work done would be transmitted from the gas to the reservoir in the course of the gas achieving thermal equilibrium with its surroundings during and after the compression. According to the ideal gas equation, if the compression of one mol of the gas were conducted reversibly (*i.e.*, sufficiently slowly that the gas would remain in virtual thermal equilibrium with the reservoir), a minimum amount of work $W_{rev}$ energetically equivalent to $RT\ln(V_1/V_2)$ of gravitational potential energy would be required to compress the gas, and, according to the First Law, an equivalent amount of thermal energy $Q_{rev}$ would be transmitted from the gas to the reservoir during the compression. In the course of this conversion of energy of the highest quality into the ambient thermal energy, the gas would be endowed with a potential for conversion of ambient thermal energy into an amount of work equivalent to that done in compressing it, as is obvious from the specifications that the temperature of the reser-voir be constant and that the compression be conducted reversibly. In other words, if one mol of the gas were compressed isothermally (reversibly) from $V_1$ to $V_2$, an amount $RT\ln(V_1/V_2)$ of actual energy equivalent in quality to gravitational potential energy would be converted into an amount $RT\ln(V_1/V_2)$ of ambient thermal energy *plus* an amount $RT\ln(V_1/V_2)$ of an energy of the gas which is ordinarily ignored when encountered in this context, but which obviously must be viewed as being free energy. If we were to consider this free energy to be actual energy, the First Law of Thermodynamics would of course appear to be very seriously in error. That it is not in error is evident from the fact that, if the free energy were not used for reversing the degradation of the nonthermal actual energy into ambient thermal energy, it would simply disappear, never to reappear. Thus, if the compressed gas were allowed to expand from $V_2$ to $V_1$ without doing work on the surroundings, as would be the case if the massless and frictionless piston holding the gas in the compressed state (Figure 1) were suddenly released, it would do so rapidly without a net change in actual energy of any kind in either the gas or its surroundings [2].

If free energy is not actually energy, what, then, is it? Since in the course of the compression phase of the above cyclic process it would become dissociated from the actual energies $W$ and $Q$, it appears in this case to be an available work potential that is free of actual energy. The facts (*i*) that complete recovery of the reversibly imparted free energy could be achieved only if the compressed gas were allowed to expand and do work at an infinitesimal rate and (*ii*) that, if the gas were allowed to expand without doing work, it would do so at the maximum possible rate with complete loss of the free energy suggest that free energy expenditure is what makes things happen within finite amounts of time. Accordingly, by combining the well-established equations of steady-state kinetics with those of classical equilibrium thermodynamics, it is possible to show quantitatively for stationary-state (apparent-equilibrium) reactions that the faster the reaction, the greater the amount of free energy consumed [3].

To see clearly the nature of the above free energy, we must consider the properties of an ideal gas in respect to actual energy. Gases in molar amounts at ordinary temperatures generally consist of very large numbers of very small particles that are moving randomly at very high speeds. They tend to occupy uniformly the entire space accessible to them and through collisions of the particles exert pressure on whatever confines them, thereby tending to increase the accessible space. The particles of an ideal gas are assumed to collide with perfect elasticity and to have volumes that are negligibly small in relation to the space available for their translation. In addition, the particles are assumed (*i*) to be devoid of attractive and repulsive forces, (*ii*) to be free of influence by fields of force such as the gravitational field, and (*iii*) to possess per mol a total amount of actual energy $E$ which depends only on the temperature. According to these assumptions, the actual energy of an ideal gas consists exclusively of thermal energy and differences in this energy on a per mol basis between states at the same temperature are zero. Thus, in the above considerations, $\Delta E$ for the gas was assumed to be zero.

We may now inquire as to the nature and source of the thermal energy that would appear in an ideal gas as it is compressed. The thermal energy must come from kinetic energy transmitted by the piston to the gas particles as the piston is moved toward the particles, this despite the fact that an amount of thermal energy $Q_{rev}$ would be imparted even if the piston were moved extremely (infinitely) slowly. The faster the piston is moved the greater the amount of energy required and thus the greater $W$ and $Q$ must be. This implies that the temperature would be determined exclusively by the translational kinetic energies of the particles and that in the course of thermal equilibration with the reservoir, the kinetic energies on average would return to their initial values, leaving only a diminished volume and a consequent augmentation of the pressure to account for the deposition of free energy in the gas. Accordingly, experimental observations on real gases suggest that allowing an ideal gas to expand without doing work (*i.e.*, without the particles expending kinetic energy) would not affect its temperature [2].

According to the ideal gas equation, $PV$ would not change in the course of an isothermal change in the state of such a gas on a per mol basis. This being the case, $RT\ln(V_1/V_2)$ for an isothermal compression would be equivalent to $RT\ln(P_2/P_1)$ and thus for such a compression one could attribute the consequent increase in free energy either to the decrease in the volume or to the increase in the pressure. That it would be appropriate to attribute the increase only to the increase in pressure is evident from the facts (*i*) that driving forces for change are determined only by intensive variables [4] and (*ii*) that pressure is the only intensive variable of the system under consideration.

It may be noted that, since there would be no difference in actual energy between the states, the increase in free energy can also be regarded as being due to an increase in the density of the translational kinetic energy or in the degree to which the translational motions of the gas particles are constrained, in which cases one can readily see that the free energy would possess an attribute of potential energy. In addition, since for an isothermal change the pressure and concentration would vary in direct proportion to one another, the free energy can also be regarded as being an energy of concentration. In fact doing so is preferable for the purposes of this study because differences in this kind of free energy between the reactants and products of biochemical reactions are ordinarily determined according to differences in concentrations. As will be pointed out below, the nature of this 'free energy of concentration' is fundamentally the same for a system in which only chemical work is possible as for the system here (Figure 1) in which only mechanical work is assumed possible.

**Free Energy in Relation to Entropy.** Since free energy is not actually energy from the standpoint of the First Law and can disappear without a trace when of the concentration variety, it is necessary to account for its consumption in terms of the actual energies. As indicated above, the actual energies $W$ and $Q$ are variable quantities that depend on how fast the compression or expansion of the gas is achieved. Although $W_{rev}$ for an isothermal (reversible) compression or expansion can be determined from the ideal gas equation and $Q_{rev}$ for such a process can be determined from $W_{rev}$ by invoking the First Law, neither $W_{rev}$ nor $Q_{rev}$ is a difference in a property, or state function, of the gas. $W_{rev}$ for such a process can be determined from the ideal gas equation because, for conditions of constant $T$, the state variables consist only of the intensive property $P$ and the extensive property $V$ which correspond to one another such that $PV$ defines an energy having dimensions of mechanical work (*i.e.*, dimensions of force × distance).

As is evident from its exclusive dependence on pressure and volume, free energy of concentration is a property of an ideal gas. However, differences in this property between states can be calculated only if the states have the same temperature. Although the free energy otherwise has the advantage that it is clearly an actual property of the gas, differences in this property between states at the same temperature are ordinarily reckoned in terms of the thermal energy that would be transmitted between the gas and its surroundings if the changes giving rise to the differences were to occur reversibly. The achievement of this requires an additional ex-tensive state property that corresponds to the intensive state property $T$ in such a way as to define an energy having dimensions of thermal energy when multiplied by $T$. In addition, since an isothermal change in the free energy would not involve a net change in the actual energy of the gas and is assumed not to involve a net change in the temperature of the surroundings, this extensive state property must be limited to thermal energy that is produced or consumed at constant temperature. In other words, (*i*) as for any other kind of energy, we must have a capacity factor for thermal energy that corresponds to the intensity factor for this kind of energy and (*ii*) the capacity factor in this particular case must be limited to thermal energy that either does not contribute to temperature or is otherwise hidden with respect to temperature.

The problem of recognizing and meeting this need was solved early in the history of thermodynamics research by Clausius [5], who noted that, despite the fact of the amount of thermal energy $Q_{rev}$ being located in the surroundings rather than in the gas, $Q_{rev}/T$ for an isothermal change in the state of an ideal gas corresponds to a change in a property of the gas. Clausius named this property 'entropy' and represented it with the symbol $S$. With respect to gases and other chemical substances, the symbol $S$ is now ordinarily used to express entropy as an intensive quantity having the same dimensions as $R$, the universal gas constant (*i.e.*, dimensions of energy per mol per degree Kelvin). As a result, $nST$ defines an amount of energy just as does $nRT$ of the ideal gas equation, $n$ being the number of mols and $nS$ the extensive component of the energy. However, whereas $nRT$ refers to an amount of pressure-volume work energy, $nST$ refers to an amount of thermal energy that is somehow hidden with respect to temperature. Of course if the thermal energy of which entropy is the capacity factor were not hidden with respect to temperature, the appropriate comparable capacity factor would be the molar heat capacity of the substance.

The molar heat capacity of a substance refers to the amount of thermal energy required to raise the temperature of one mol of the substance by one degree Kelvin. The heat capacity of a gas can be determined unequivocally only under conditions either of constant pressure or of constant volume, in which cases the molar heat capacities are denoted by $C_P$ and $C_V$, respectively. According to the classical kinetic theory of gases, $C_P$ and $C_V$ for an ideal gas would be constants having the values $\frac{5}{2}R$ and $\frac{3}{2}R$, respectively, regardless of the magnitudes of the state properties temperature, pressure, and volume at which they are determined [6]. $C_P$ exceeds $C_V$ by $R$ because an amount of pressure-volume work equivalent to $R$ would be done on the surroundings if the heat capacity were determined at constant pressure. Thus, only $C_V$ is a heat capacity that refers exclusively to energy possessed by the gas.

As one might expect from its relationship to free energy of concentration, entropy in general simply appears, never to disappear, as free energy vanishes. As a result of the fact that its total amount invariably increases, entropy is often said not to be conserved, which, of course, is true in the sense that the amount is not a constant. Since the increase is determined by the amount and quality of the actual energy that invariably appears as high-quality actual energy and its associated free energy disappear, it seems more appropriate and instructive to acknowledge instead that it is the free energy (quality) of actual energy, or high-quality actual energy itself, that is not conserved. Nevertheless, entropy has been universally adopted as the principal index of free energy consumption, this despite the facts that the quality of an energy is determined by the magnitude of its intensive component [4] and that entropy, being, like volume, an extensive component of an energy, is consequently relatively incomprehensible as an indicator of energy degradation. However, as will be outlined below, entropy has an important advantage over free energy in this respect in that it can be and has been universalized such as to make possible a proper (practical) assessment of the quality of thermal energy.

In this regard it is important to note that the free energy of the intermediary thermal energy of a thermodynamic work process mediated by an ideal gas would be subject to loss by two means depending on whether the thermal energy moves spontaneously down a gradient of pressure by convection, in which case the free energy that is subject to loss would be the above-described free energy of concentration, or down a gradient of temperature by conduction and/or radiation. If entropy is to be a universal capacity factor, it must be capable of serving as the capacity factor for free energy that is subject to loss by both these means. In the case of convection, the intensive and extensive components of the energy are pressure and volume and the extensive component can be readily expressed in terms of entropy. Thus, for an isothermal (reversible) compression of an ideal gas, the increase in free energy of concentration on a per mol basis can be represented either by $-RT\ln(V_2/V_1)$ or by $-T\Delta S$, $\Delta S$ being equivalent to $R\ln(V_2/V_1)$, a negative quantity in the case of compression, indicating a decrease in the entropy of the gas. Of course the reason why the entropy of the gas would decrease in this case is that the thermal energy on which the entropy is based would be located in the surroundings rather than in the gas. Entropy is universalized by using this volume-dependent entropy in a Carnot cycle to define an entropy in terms of the thermal energy possessed by the gas, in which case the free energy that is subject to loss would have temperature and heat capacity as its intensive and extensive components. Since both $C_V$ and $C_P$ would be constants, any changes in this 'derivative' entropy must necessarily be expressed in terms of changes in the temperature and thermal energy of the gas.

As is evident from the above-described properties of free energy of concentration, the volume-dependent entropy of an ideal gas could change through either compression or expansion without there being a change in the actual energy of the gas and, in the case of expansion, could change in the absence of a net change in actual energy of either the gas or its surroundings. Entropy being a state property, it is also evident that the volume-dependent entropy would not depend on the rate of change. In contrast, the temperature-dependent entropy, being based on thermal energy possessed by the gas, would depend greatly on the rate of change, a fact which can be readily seen by considering irreversibility in respect to an adiabatic cycle of compression and expansion of an ideal gas (see **APPENDIX**).

In consequence of the different locations of the actual energies on which the volume- and temperature-dependent entropies are based, one kind could increase while the other decreases if the gas were to undergo a change in temperature and thermal energy as well as in volume. In view of the marked differences between the two entropies, it seems highly desirable to distinguish between them on a regular basis more or less as done originally by Gurney [7]. In making this distinction in what follows, the kind of entropy that depends only on the volume shall be referred to as entropy of concentration to correspond with free energy of concentration. That this is appropriate can be seen by noting that the concentration of a given amount of an ideal gas would vary inversely as the volume regardless of the temperature and pressure and that concentration can therefore be substituted for volume in the expression $R\ln(V_2/V_1)$ for a difference in the volume-dependent entropy on a per mol basis simply by changing a sign. The kind of entropy that would depend only on the temperature and thermal energy of the gas shall be referred to as characteristic entropy, and the corresponding kind of free energy shall be referred to as characteristic free energy. It is important to note that these 'concentration' and 'characteristic' kinds of entropy and free energy are very similar to but not entirely identical with the 'cratic' and 'unitary' kinds defined by Gurney.

In contrast to changes in free energy of concentration, changes in characteristic free energy are invariably accompanied by equivalent changes in the energies recognized by the First Law. Since, as noted above, the only actual (First-Law) energy of an ideal gas would be undirected translational kinetic energy, the characteristic free energy of such a gas must necessarily be of the above-mentioned, 'potential' kind possessed by thermal energy (*i.e.*, that recoverable through work only at the expense of a decrease in the temperature of the energy as its work potential is being realized). In view of the prediction of the ideal gas equation that the size of a change in the free energy of concentration due to a compression or an expansion between two states of concentration at the same temperature can be determined on a per mol basis according to $RT\ln(V_2/V_1)$ or $RT\ln(P_2/P_1)$ and thus would be directly proportional to temperature, it is evident that changes in free energy of concentration, although not involving net changes in the characteristic free energy of the gas, would require that the gas possess characteristic free energy if the changes are to be finite, the characteristic free energy being that associated with the thermal energy at the temperature of the two states. Of course the higher the temperature, the greater the change in free energy of concentration for a given change in concentration of the particles possessing the thermal energy and free energy.

Changes in the characteristic free energy are ordinarily reckoned in terms of work done on or by the gas during its compression or expansion under adiabatic conditions (*i.e.*, under conditions in which transmissions of thermal energy between the gas and its surroundings cannot occur). If this is to be achieved, one must have knowledge of the capacity factor for thermal energy that determines temperature and refers only to energy possessed by the gas. As noted above, $C_V$, the molar heat capacity determined under conditions of constant volume, meets this requirement and, for an ideal gas, would not depend on the magnitudes of the state properties at which it is determined. This being the case, the total energy for one mol of such a gas would be given by $C_V T$ and finite changes in total energy on a per mol basis by $C_V \Delta T$, $C_V$ being a constant. Since reversible processes do not expend free energy, $C_V \Delta T$ for any reversible adiabatic compression or expansion would be equivalent to the change in characteristic free energy as well as to the change in total energy. Thus, if the quality of thermal energy were judged on the basis of the predicted changes in the characteristic free energy of an ideal gas, it would appear to be equivalent to that of nonthermal actual energy. However, as outlined below, a proper assessment of the quality of thermal energy in terms of changes in the properties of a gas can be achieved only by means of the Carnot cycle.

**The Carnot Cycle.** Owing to the extensive use of gas heat engines to exploit the work potentials of natural sources of nonthermal actual energy, an important concern of thermodynamics is the availability (quality) of the work potential of thermal energy imparted to a gas in the course of the gas undergoing a cycle of expansion and compression. Since in a strictly adiabatic or strictly isothermal cyclic process, a gas cannot do more work than is done on it, net conversion of thermal energy into nonthermal actual energy is impossible in these cyclic processes. As pointed out originally by Carnot [8], such a conversion can be achieved in a cyclic process only by combining isothermal and adiabatic processes in an alternating sequence in which the gas is allowed to undergo isothermal expansion with uptake of thermal energy from a reservoir at a relatively high temperature, followed by adiabatic expansion and isothermal compression with transmission of a smaller amount of thermal energy from the gas to a reservoir at a relatively low temperature. Net conversion is achieved exclusively as a result of the fact that a gas at a relatively high temperature can do more work through isothermal expansion than is required for isothermal compression by the same factor at a relatively low temperature.

If all four steps of the Carnot cycle were conducted reversibly, none of the characteristic free energy of the thermal energy absorbed isothermally by the gas at the relatively high temperature and converted through work into the kind possessed by nonthermal actual energy would be consumed. However, complete conversion of the free energy would be possible only if the gas could undergo infinite expansion in the adiabatic expansion phase of the cycle and the relatively cold gas and reservoir were at a temperature of absolute zero. Therefore, since available machines are of limited size, and since the thermal sinks ordinarily available have temperatures much higher than absolute zero, complete or nearly complete conversion of thermal energy into energy equivalent in quality to gravitational potential energy is far from being practical. Of course this is the basis for the limited potential availability of the work potential of thermal energy.

As is evident from the Carnot cycle and the above-noted fact that the quality of an energy is determined by the magnitude of its intensive component, the practical quality of thermal energy is higher as the temperature is higher. Determination of the quality by means of the Carnot cycle requires knowledge of the relationship between temperature and volume in the adiabatic steps. The required information is obtained by defining a characteristic entropy in terms of entropy of concentration, which, as noted above, would be a function only of volume and concentration for an ideal gas. Since entropy of concentration is a capacity factor for thermal energy that is hidden with respect to temperature, this characteristic entropy must be scaled to thermal energy transmitted to or from the gas at constant temperature, a fact which is relevant to the above-mentioned need to know the relationship between temperature and volume in the adiabatic steps. The thermal energy to be evaluated as to quality is that transmitted to the gas and converted into work isothermally at the relatively high temperature. Since entropy refers to thermal energy, it is the 'thermal fraction' of this thermal energy (*i.e.*, the fraction transmitted isothermally to the surroundings at the relatively low temperature) to which changes in the characteristic entropy of the adiabatic steps must be scaled. Of course the isothermal transmissions are assumed to be possible as a result of the reservoirs being sufficiently large to be capable of yielding and accommodating thermal energy without undergoing a change in temperature.

Owing to the fact that it would not be possible for the temperature and thermal energy of an ideal gas to change independently of one another, a change in the characteristic entropy can be expressed properly in terms of the temperature and thermal energy only by means of the differential equation $dS_{char} = dE/T$ or its equivalent. Since $dE = C_V dT$ and $C_V$ would be a constant, this equation can be expressed in the forms $dS_{char} = C_V dT/T = C_V d\ln T$ and integrated between two specific temperatures to yield $\Delta S_{char} = C_V \ln(T_2/T_1)$. The comparable expression $R\ln(V_2/V_1)$ for a finite change in the entropy of concentration on a per mol basis is obtained essentially in the same manner. In this case, however, it is the pressure and volume that would be incapable of changing independently of one another. In consequence of this and of the fact that the volume must change if the entropy is to change, a change in $S_{conc}$ can be expressed properly in terms of volume and pressure only by means of the differential equation $dS_{conc} = PdV/T$, which can be modified by substituting $RT/V$ for $P$ to obtain the equation in the integrable form $dS_{conc} = R\, d\ln V$. Of course the scaling of $\Delta S_{char}$ to $\Delta S_{conc}$ in the adiabatic steps is accomplished through the facts (*i*) that a net change in entropy would not be possible in a reversible adiabatic process and (*ii*) that, in consequence, the ratios of the initial and final temperatures and volumes in the adiabatic steps of a reversible Carnot cycle would be constrained to agree with one another according to the relationship: $T_2/T_1 = (V_1/V_2)^{R/C_V}$.

That $\Delta S_{char}$ for the adiabatic steps is based on the thermal energy transmitted isothermally to the surroundings at the relatively low temperature is evident from the facts (*i*) that the maximum efficiency $\eta_{max}$ of a Carnot heat engine is equivalent to $1 - T_{low}/T_{high}$ and (*ii*) that $\Delta S_{char}$ for the adiabatic steps of the reversible cycle is therefore equivalent to $\pm C_V \ln(1 - \eta_{max})$, the positive and negative signs referring to the expansion and compression steps, respectively. By noting that $\eta_{max}$ refers to the available (work) fraction of the thermal energy absorbed at $T_{high}$, one can readily see that $\Delta S_{char}$ refers to the unavailable (thermal) fraction. For any given finite value of $T_{high}$, the unavailable fraction is a linear function of $T_{low}$ and decreases to zero as $T_{low}$ approaches zero.

The Carnot cycle is very important in that it universalizes entropy and thereby makes it possible to evaluate the practical quality of thermal energy and to demonstrate by theoretical means for thermodynamic work processes in general that, if such a process is to occur at a finite rate, free energy must be expended and thermal energy must be produced in an amount equivalent to the amount of free energy consumed, a fact which can be readily demonstrated by considering irreversibility in respect to an adiabatic cycle of compression and expansion of an ideal gas (see **APPENDIX**). In any particular case, the fundamental process giving rise to the finite rate would be the net conversion of relatively high-quality actual energy, both thermal and nonthermal, into ambient thermal energy, the spontaneous and unidirectional nature of which is the basis for the Second Law of Thermodynamics, which, unlike the First Law, acknowledges the existence of free energy and says in effect that if thermodynamic work is to be done at a finite rate, free energy must be expended. Also unlike the First Law, the Second Law, owing to the fact that the individually mobile particulate constituents of macroscopic amounts of matter at finite temperatures vary widely as to translational kinetic energy, is a statistical law appropriate for application only to macroscopic phenomena. This means that the Second Law is obeyed only on average over time in processes at the microscopic level and thus that conversions of ambient thermal energy into nonthermal actual energy in chemically active substances can occur at the molecular level. Of significance in this regard is the fact that translational thermal energy at the molecular level is kinetic energy of the directed variety, a consequence of which is that no distinction can be made between this form of actual energy and nonthermal actual energy at the molecular level. Accordingly, energy transfer at the molecular level occurs without expenditure of free energy, and irreversibility, like temperature and pressure in respect to a gas, is a concept applicable only to macroscopic phenomena.

**Real Gases.** As is well known, the heat capacities of real gases increase with increase of temperature and correspond closely to those predicted for an ideal gas over wide ranges of temperature and pressure only for monatomic gases [9]. The temperature dependence is due in large part to the fact that the translational and radiant forms of thermal energy are capable of undergoing interconversion with actual energy associated with quantized motions within molecules. The energies of these 'intramolecular' motions are reckoned in terms of characteristic entropy and thus appear to be generally viewed as being of a thermal nature, this despite the facts (*i*) that the energies of some of the motions undergo oscillatory interconversions with attractive and repulsive potential energies and (*ii*) that the motional energies, being intramolecular, seem best viewed as being directed (nonrandom) kinetic energies and thus of a nonthermal nature. On the other hand, since radiant energy is not confined to molecules and is ordinarily unrestricted as to direction of emission in a gas, any radiant energy emitted as a result of the motions would clearly qualify as thermal energy.

In accord with their being nonthermal, the energies associated with the intramolecular motions apparently do not contribute to temperature. Such is consistent with the facts (*i*) that the absolute and thermodynamic scales of temperature are based on the properties of an ideal gas and (*ii*) that an ideal gas is assumed to possess actual energy only of the translational kinetic kind. It may be noted that, owing to the discontinuous nature of the intramolecular motions in respect to change of temperature, such must be assumed for validity of the widely accepted, sweeping generalization concerning the reversible Carnot cycle that the ratio of the amount of thermal energy absorbed isothermally at $T_{high}$ to the amount rejected isothermally at $T_{low}$ would have the same value regardless of the nature of the working substance.

Despite their apparent nonthermal nature, the energies associated with the intramolecular motions appear to be recoverable only as thermal energy at the temperature of their production and thus appear to differ markedly from the kind of nonthermal energy that is capable of converting spontaneously into thermal energy of quality exceeding that of the ambient thermal energy. Accordingly, their elicitation is associated with diminution of molecular stability [10] and might be expected thereby more likely to diminish than to enhance chemical potential energy of the kind that possesses characteristic free energy. This kind of energy shall henceforth be referred to as 'characteristic chemical potential energy', a distinction made necessary by the fact that free energy of concentration is ordinarily treated as if it were chemical potential energy, a practice which of course is valid when properly used but which ignores the intrinsically mechanical nature of the concentration-dependent free energy and tends to elicit confusion as to the nature of characteristic chemical potential energy, particularly when used in reference to an ideal gas.

Although, as noted above, the energies associated with the intramolecular motions also differ appreciably from what is ordinarily considered to be thermal energy, in view of the current practice of accounting for them appropriately in terms of characteristic entropy and of entropy being the capacity factor for thermal energy that is hidden with respect to temperature, it seems appropriate and best for practical purposes to view the energies as being latent forms of thermal energy of quality determined by the temperature at which they could be reversibly produced. By making this distinction we imply that thermal energy that is not latent refers to the kind that determines temperature. Since what has been referred to above as the radiant form of thermal energy is actually electromagnetic energy having frequency $\nu$ as its intensive component and $nh$ as its extensive component, $n$ being any whole positive number and $h$ Planck's constant, and can be said to be thermal energy only in the sense that it is capable of transmitting thermal energy and to have a temperature only in virtue of the fact that it has a certain distribution as to quality and concentration of photons of energy $h\nu$ as given by Planck's Law of Heat Radiation when in equilibrium with matter at a particular temperature [11], this 'nonlatent' thermal energy may be considered to consist exclu-sively of the translational kinetic kind. However, since electromagnetic energy is readily detectable and transmissible as such, when viewed as being thermal energy it must in some sense also be viewed as being a nonlatent variety, particularly in respect to transmission of thermal energy.

The intramolecular motions include rotations of entire molecules and various rotations, librations, and vibrations of molecular constituents, all of which are known to be quantized through the occurrence of temperature-specific changes in heat capacity and in absorption and emission of characteristic radiant energy [10, 12]. Of course the possibilities for these kinds of motion are greater, the greater the complexity of the molecules and the weaker and more flexible the bonds between the constituent atoms. In view of the fact that the intramolecular motions can be elicited through inelastic collisions between molecules, these motions must be capable of converting into the translational motions that result in the collisions and, under conditions of constant temperature, must tend to be at equilibrium with the translational motions; otherwise, contrary to Planck's Law and the Zeroth Law of Thermodynamics, there would likely be a temperature differential between the translational and radiant forms of thermal energy at equilibrium.

Due to the existence of net attractive forces between the individually mobile particulate constituents of macroscopic amounts of real matter, thermal energy can also disappear and appear with increase and decrease of temperature through dissociation and association reactions that increase and decrease the number of particles whose motions contribute to and thereby determine temperature. This phenomenon can be explained on the basis of the very successful prediction of the equipartition principle of the classical kinetic theory of gases that, at any particular temperature, individually mobile particles differing as to mass, composition, and other properties will possess on average the same amount of translational kinetic energy, the amount being equivalent to $\frac{1}{2}m\bar{v}^2$, $m$ being the mass and $\bar{v}^2$ the mean square velocity [9, 12, 13]. According to this prediction, if the molecules of a gas at a particular temperature were sufficiently attracted to one another that some of the molecules could expend attractive binding energy by binding to one another at that temperature, increasing the temperature would result in mechanically induced dissociations of bound molecules and in a portion of the thermal energy added to the gas for the purpose of increasing its temperature being expended to elevate the translational kinetic energies of newly formed particles to average values consistent with the existing temperature. Since the formation of the additional particles would be accompanied by the appearance of attractive forces and thus also of attractive binding energy, it seems appropriate to view the dissociations of the attractively bound particles as constituting conversions of translational kinetic (mechanical) energy into attractive binding energy. Also, since it is unlikely that attractive binding energy thus generated could be recovered as an energy of quality higher than the thermal energy required for its reversible production, and since the attractive forces would extend beyond the bounds of the particles possessing them, it seems appropriate to view the attractive binding energy as being an extramolecular kind of latent thermal energy of quality determined by the temperature at which it could be reversibly produced. As will be pointed out below, one could reasonably expect the attractive binding energy to have the quality of characteristic chemical potential energy only if the gas were supercooled with respect to thermochemical equilibrium.

Since associations of molecules through mutual attractions are likely to result in some hindrance of intramolecular motions, changes in latent thermal energy of the extramolecular kind at temperatures sufficiently high for elicitation of intramolecular motions are likely to be accompanied by changes in latent thermal energy of the intramolecular kind in the same direction. In contrast to the intramolecular kind, the extramolecular kind can undergo change in response to change not only of temperature, but also of pressure and thus of concentration. Both kinds can change under conditions of constant temperature and pressure in chemical reactions and in first-order phase transitions, and, since both are based on system energies having the quality of thermal energy, changes in their amounts under these conditions are actually and best accounted for in terms of characteristic entropy.

As indicated above, the prediction of the classical kinetic theory of gases that $C_V$ for an ideal gas would be a true constant implies that the amount of thermal energy possessed by a given amount of such a gas would be directly proportional to temperature. This in turn implies that an ideal gas can be considered to provide linear absolute scales for actual energy and free energy as well as for temperature. Although based on predicted properties of a fictitious gas that differs appreciably from real gases in that its particulate constituents possess only translational kinetic energy, these scales are very important in that they are commonly used with remarkable success as a framework for characterization of the thermodynamic properties of all real substances, a fact which accords with the universality of the universal gas constant $R$. As will in effect be suggested below, the widespread success of the ideal gas model is likely due in large part to the above-noted prediction of the equipartition principle being applicable to liquids and solids as well as to gases and to temperature in liquids and solids being determined by the average kinetic energies of individually mobile particles consisting of clusters of attractively bound molecules, the average size of which tends to increase with decrease of temperature.

**THERMODYNAMICS OF LIQUIDS AND SOLIDS**

All naturally occurring gases possess net attractive forces and in consequence undergo condensations to form liquids and solids as temperature is decreased. An important but rarely asked question is: What determines temperature in these condensed phases? This question must be asked and answered correctly if we are to understand how it is possible that equilibrium differences in concentration between the reactants and products of a chemical reaction conducted in solution can serve to measure differences in characteristic chemical potential energy between the reactants and products at various equilibrium temperatures.

Since the particulate constituents of liquids can flow and thus must be sufficiently free to undergo translational motions, we can attribute temperature in this case to the same kind of motions that determine temperature in gases. Since the particulate constituents are held in the liquid state by attractive (cohesive) forces possessed by the particles, one might expect the translational motions of the particles to be hindered in respect to intensity of translational motions and thus in respect to ability to contribute to and thereby determine temperature. However, in view of (*i*) the likelihood that temperature is determined in both liquids and gases by the average intensity of the *actual* translational motions of the constituent individually mobile particles and of (*ii*) the above-noted prediction of the equipartition principle that, at any particular temperature, the individually mobile particulate constituents of all gases will possess on average over time the same amount of translational kinetic energy, it seems likely that the average translational kinetic energy possessed by the particulate constituents of any liquid would tend to be the same as that for any gas if the liquid and gas were at the same temperature. If such were not the case, it would be very difficult to understand, among other things, the physical bases for the latent thermal energies and for the common observation that the mechanical thermodynamic properties of substances in dilute (ideal) solution do not differ appreciably from those expected of an ideal gas despite the individual molecules of the substances differing greatly as to such properties as chemical composition, size, net charge, and affinity for the solvent.

As is well known, condensations of gases into liquids and of the liquids into solids under conditions of constant pressure can occur very nearly reversibly at constant temperature with productions of amounts of thermal energy that greatly exceed the amounts of pressure-volume work done simultaneously. It is important to inquire as to where the excess thermal energy comes from in these processes. Since the condensations occur as a result of there being attractive forces between the particulate constituents, it must necessarily come primarily from conversions of attractive binding energy into thermal energy as the particles associate in the course of the condensations. In view of this and of the likelihood that the average translational kinetic energies of the individually mobile particulate constituents of at least the liquid and gas phases of any substance would be the same if the phases were at the same temperature, it seems likely that such conversions occur through associations of particles to form larger and thus fewer particles possessing on average the same amount of translational kinetic energy as the particles undergoing the associations. Thus it seems appropriate to view the attractive binding energy that undergoes conversion into thermal energy in phase transitions of the above sort to be latent thermal energy of the above-described extramolecular kind.

As is particularly well known, thermally induced transitions of water from solid to liquid and from liquid to vapor at normal (atmospheric) pressures involve large conversions of thermal energy into latent thermal energy. It is also well known that some of the latent thermal energy is of the intramolecular kind and that the conversions are capable of occurring at constant temperature with little consumption of free energy when the nonlatent thermal energy derives from ambient thermal energy [14]. By linking reversible production of latent thermal energy in these processes to isothermal (reversible) compression of an ideal gas, we can readily see that both the intramolecular and the extramolecular kinds of latent thermal energy must in fact be viewed as being equivalent in quality to ambient thermal energy at the temperature of their reversible production. Consider, for example, a Figure 1 system in which the temperature of the thermal reservoir is at the melting temperature of ice and is maintained constant solely by interconversions between ice and liquid water at constant external pressure. If the gas in such a system were compressed reversibly, thermal energy would be produced isothermally in the gas and consumed isothermally in the reservoir through reversible conversion of ice into liquid water. Since all of the free energy of the nonthermal actual energy imparted to the gas would be retained by the gas and could be conserved only through reversal of the above process, it would be necessary to consider any latent thermal energy produced as a result of translational thermal energy from the gas converting ice into liquid water to be equivalent in quality to ambient thermal energy at the melting temperature of the ice.

The above observations being correct, one could reasonably expect the number of individually mobile particles in a liquid to increase as the temperature is increased. Since the increase would necessarily occur through dissociations of attractively bound particles, one could reasonably expect the heat capacities of liquids to be augmented by conversions of thermal energy into attractive binding energy as the temperatures of the liquids are increased for the purpose of measuring the heat capacities. That these expectations are consistent with what is observed experimentally in this regard is particularly clear in the case of water.

Owing to the molecules of water as compared to those of most other common solvents having particularly strong tendencies to associate with one another through their capacities to serve both as a double donor and a double acceptor of hydrogen bonds, the energy changes in the gas ⇌ liquid and liquid ⇌ solid transitions of water are particularly large. The molar heat capacity of the liquid at constant atmospheric pressure is also particularly large and is close to twice that of the solid near the normal freezing point and to twice that of the vapor near the normal boiling point [15]. The above-noted expectations are clearly consistent with these experimental findings and with numerous experimental and theoretical observations suggesting that the particulate constituents of liquid water consist largely of labile clusters of molecules, the concentration and average size of which depend on temperature and pressure [15-18].

For there to be consistency with the above observations, it is necessary to suppose that, for solids, temperature and to a large extent also heat capacity are determined by undirected translational (vibrational) motions of clusters of molecules (atoms in the case of monatomic solids) about mean fixed positions and that the average size of these clusters tends to increase with decrease of temperature, the increase in average size being accompanied by conversion of kinetic and potential (attractive + repulsive) vibrational energy into ambient thermal energy. In view of the fact that the average vibrational frequency of the clusters would decrease as the average size (mass) of the clusters increases and of experimental observations suggesting that the heat capacities of all naturally occurring substances tend to approach zero as temperature approaches zero, in considering this to be the case for a large chunk of a solid maintained close to thermochemical equilibrium, it would seem necessary to assume that the number and average vibrational frequency of the clusters would decrease to unity and zero, respectively, as the temperature of the solid is decreased to zero. This and the further assumption that the atoms or molecules of the individual clusters vibrate coherently and consequently emit photons as clusters of indistinguishable photons being correct, one could think of the clusters as being Bose-Einstein condensates of a sort and of the solid as being a Bose-Einstein ideal gas in which the number of particles is not conserved more or less as the Bose-Einstein quantum statistical method of deriving Planck's Law of Heat Radiation predicts for distinguishable clusters of indistinguishable photons in equilibrium with a blackbody (see below). Doing so would be consistent with the generally accepted Bose-Einstein quantum gas, 'quasiparticle' (phonon) interpretation of lattice vibrations in crystalline solids [11, 19, 20] and thus with the apparent wave-particle duality of matter in this case. Doing so would also be consistent with experimental observations [21, 22] indicating that water ice at subfreezing temperatures sublimates (evaporates) in the form of clusters of molecules.

In 1913 prior to the general acceptance of the quantum hypothesis demanded by the extremely close agreement between experimental observations and the equation now known as Planck's Law of Heat Radiation, Benedicks [23] presented an 'agglomeration hypothesis' which appears to be generally consistent with the above notions concerning what determines temperature and heat capacity in solids. Thus, Benedicks pointed out that Planck's Law can be derived on the basis of the assumptions (*i*) that the atoms or molecules of solids coalesce through cohesive forces to form clusters of increasing size as temperature is decreased and (*ii*) that the thermal energy partitions among the clusters in accordance with the equipartition principle. He also pointed to the possibility of explaining the observed independence of the frequency spectrum of cavity (equilibrium blackbody) radiation on the nature of the solid by taking into account the facts (*i*) that the average mass of the clusters at a particular temperature would be greater as the cohesive force between the constituent atoms or molecules is greater and (*ii*) that, since the average vibrational frequency of the clusters would be lower as the average mass of the clusters is greater and higher as the cohesive force between the clusters is greater, differences among solids as to cohesive force would tend to cancel in respect to the vibrational frequency spectrum of the clusters at a particular temperature and thus also in respect to the frequency spectrum of the radiant energy emitted and absorbed by the clusters at that temperature.

In 1915 A. H. Compton [24] tested the agglomeration hypothesis as to utility in accounting for experimentally observed relationships between temperature and heat capacity in simple solids. Using Maxwell's Distribution Law in conjunction with the agglomeration hypothesis, he came up with a very simple equation which he judged to be at least equally as accurate as the much more complex but now generally accepted equation developed earlier by Debye [25] on the basis of the quantum hypothesis using arbitrarily the assumption that vibrational frequencies at low temperatures are limited to the low values that one might expect to be characteristic of large clusters of atoms. Compton [26] also compared the agglomeration and quantum hypotheses as to utility in accounting for observed inverse relationships between temperature and thermal conductivity in crystalline solids and found the agglomeration hypothesis to be clearly superior in this case.

Despite its remarkable successes, the agglomeration hypothesis was largely abandoned, presumably as a result of Compton turning his attention to his well-known studies on interactions between X-rays and matter, which resulted in the general acceptance of the quantum hypothesis, and to the textbook view [27] that if the agglomeration hypothesis were valid, one could expect solids to be incompressible at temperatures approaching absolute zero, a view which has persisted into modern times [28] despite the fact that it was based on the faulty notion that atoms are incompressible (hard) spheres (see [29]). Nevertheless, although not acknowledged or recognized, the main features of the agglomeration hypothesis were used in the subsequent development of quantum mechanics, a fact which is particularly evident in the case of wave mechanics. Thus, Bose-Einstein (quantum) statistics become essentially identical with Maxwell-Boltzmann (ideal-gas) statistics as temperature is increased and assumes a condensation (degeneration) of distinguishable particles (parcels) of energy into distinguishable clusters (cells) of indistinguishable particles of energy as temperature is decreased [9]. Also, de Broglie's fundamental ideas on wave-particle duality, which, in conjunction with Bose-Einstein statistics led Schrödinger immediately to his wave mechanics [30], originated with the realization that derivation of Planck's Law by considering light quanta to be an ideal gas of photons ("atoms of light") requires the assumption that blackbody radiation other than that at the extreme high-frequency end of the observable spectrum consists of agglomerations of photons that move coherently [31-33]. Since the frequency at the high-frequency end of the observable spectrum increases endlessly with increase of temperature, this assumption and the essentially identical one at the heart of Bose-Einstein statistics being correct, most if not all 'conventional photons' would be clusters of photons having the same frequency and we could imagine a rational explanation for the peculiar fact of quantum mechanics that what appear to be single photons can appear to split and interfere with themselves in a wave-like manner. However, the explanation would not be of a sort that would seem likely to account for the considerably more peculiar, closely related fact that electrons, neutrons, and other genuine particles in what seem highly likely to be genuine (indivisible) single-particle states also can appear to split and interfere with themselves in a wave-like manner (see, *e.g.*, [34, 35]).

That temperature and heat capacity in solids is actually and strongly linked to the average size of actual particles consisting of clusters of atoms or molecules is suggested by the results of numerous relatively recent studies on the thermodynamic properties of 'nanosolids' obtained by reducing normal (bulk) solids at temperatures far below their melting points to particles having diameters of a few nanometers and then lightly compacting the particles to form pellets that can be easily handled and compared with the bulk materials (for comprehensive reviews, see [36, 37]). In general, the thermodynamic properties of solids thus modified have been found to differ from the normal such as to suggest that the temperature has in effect been greatly increased. Thus, melting temperatures, at least of individual particles, are greatly decreased [38], enthalpies [39], entropies [36], heat capacities [36, 39], and vapor pressures [40] are greatly increased, and heat capacities appear not to approach zero as temperature approaches zero [36, 41]. Most of these changes either have been or can be adequately explained in terms of the associated large increases in surface area, surface attractive binding energy, and number of particles that are individually mobile under conditions of temperature where the attractive binding energy has the quality of characteristic chemical potential energy rather than of latent thermal energy (see below). The observed persistence of heat capacity in lightly compacted nanosolids at temperatures very close to zero is consistent with the theoretical findings of Jura and Pitzer [42] predicting that the heat capacities of solids consisting of unconsolidated 'nano-sized' particles will be detectably large at temperatures closely approximating absolute zero as a result of the particles being capable of undergoing thermally induced translational and rotational motions at such temperatures despite the particles possessing an abundance of attractive binding energy.

**THIRD LAW OF THERMODYNAMICS**

According to the first of two extant versions of the Third Law that are relevant here, the entropy change associated with any isothermal (reversible) process will approach zero as temperature approaches zero [43, 44]. It is important to inquire as to the nature of the entropy to which this version of the law refers. Since it concerns only processes occurring at particular temperatures, characteristic entropy of the kind that can change only as a result of a change in temperature can be immediately ruled out, thereby avoiding the uncertainty associated with the prediction of the kinetic theory of gases that a reversible decrease in the temperature of an ideal gas to zero in a Carnot cycle would be accompanied by a decrease in characteristic entropy to $-\infty$, a result necessitated by the fact that achievement of the condition $T = 0$ in the cycle would require increases in the volume and entropy of concentration to $+\infty$. On the other hand, entropy of concentration can undergo change at constant temperature and can do so without necessarily requiring a change in volume, a fact which can be readily seen by considering chemical reactions of the sort: A $\rightleftharpoons$ P + Q (*i.e.*, reactions in which there is a difference in mol number between the reactants and products in the stoichiometric equation). As will be outlined below, at least for reactions occurring in solution virtually at constant volume, a change in entropy of concentration due to a change in the number of particles in such a system maintained at constant temperature will be accompanied by a change in the characteristic entropy of the system and the thermal energy on which the characteristic entropy is based will be indistinguishable from the above-described extramolecular latent kind. That a change in entropy of concentration would conform to the above version of

*Nature of Free Energy and Entropy*

the Third Law can be seen by noting that changes in entropy of concentration at a particular temperature serve to represent changes in free energy of concentration and that, since free energy of concentration depends on the existence and density of translational kinetic energy and therefore can be expected to be zero at $T = 0$, a change in entropy of concentration would not be possible at $T = 0$.

The same can be said of entropy of mixing, the nature of which can readily be seen to be identical with that of entropy of concentration by considering a spontaneous mixing of two distinguishable ideal gases at the same temperature and initial pressure. A change in entropy of mixing in this case is given by the sum of the increases in the entropies of concentration of the individual gases as each diffuses (expands) into the space occupied by the other. According to this, entropy of mixing serves to account for the free energy of concentration that could be used for conversion of ambient thermal energy into nonthermal actual energy if there were a means by which the diffusion of the substances undergoing spontaneous mixing could be harnessed individually to do work, a possible means in the above case being an apparatus consisting of a cylinder having two chambers separated initially by two opposed semipermeable pistons, the piston nearest each gas in its pure form being permeable only to that gas. In the absence of such means, the mixing would occur without a net change in actual energy, just as would be the case if the individual gases were expanding into a vacuum.

Presumably, the principal kind of entropy to which the above version of the Third Law is ordinarily meant to apply is the characteristic entropy associated with the kinds of energy referred to above as intramolecular and extramolecular latent thermal energy. As was noted, these forms of energy do not contribute to temperature and can undergo change in processes occurring at constant temperature. Since, at any particular finite temperature, they tend to be at equilibrium with the transmissible (nonlatent) forms of thermal energy, they can be expected to disappear as temperature approaches zero if sufficient time is allowed for them to do so.

Owing to the tendency of the latent thermal energies to equilibrate with the transmissible forms, the Third Law is widely interpreted to mean that the characteristic entropies of individual substances approach zero as temperature approaches zero. Although the version thus obtained has proved to be very useful for practical purposes and is consistent with experimental observations suggesting that the heat capacities of all naturally occurring substances in macroscopic amounts maintained close to thermochemical equilibrium will decrease to zero as the temperature is decreased to zero, it is considered not to be generally correct as a result of the possibility for the characteristic entropy associated with the extramolecular kind of latent thermal energy to become 'frozen in' as the temperature is decreased. As implied above, the extramolecular kind of latent thermal energy is clearly nonthermal, is possessed by all naturally occurring substances at finite temperatures, and consists of attractive binding energy that is subject to loss upon temperature reduction through stabilization of neutralizing attractive binding interactions among the particulate constituents whose undirected motions contribute to and thereby determine temperature. In this particular case, latent thermal energy and its associated entropy can become 'frozen in' as a result of free molecules condensing during temperature reduction to form stable solids in which the molecules are not arranged as well as they could be for maximum conversion of the binding energy into the radiant and translational forms of thermal energy, which, in contrast to the binding energy, are transmissible as such and subject to direct removal through reduction of the ambient temperature. For reasons of this sort, the Third Law, as it is applied to individual substances, is generally considered to be applicable in a strict sense only to pure substances in perfect crystalline states [45].

**THERMODYNAMICS OF BIOCHEMICAL REACTIONS**

We shall be concerned here primarily with the very simple, first-order, uncatalyzed reaction:

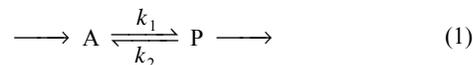

$$\longrightarrow A \underset{k_2}{\overset{k_1}{\rightleftarrows}} P \longrightarrow \qquad (1)$$

in which A and P are reactant and product in ideal solution at constant temperature and pressure, $k_1$ and $k_2$ are the forward and backward rate constants, and the arrows preceding the reactant and following the product are meant to indicate that $a$ and $p$, the concentrations of A and P, are maintained constant as a result of the reaction being in a long sequence of reactions of the sort that one might expect to find in living systems. Thus, Reaction (1) as presented is considered to be a stationary-state reaction in an open system of more or less constant volume and to have a net velocity in the direction indicated regardless of whether the equilibrium constant is favorable or unfavorable. Although the system as presented is considered to be open and therefore relevant primarily to biological systems, in most of what follows it will be considered to be closed (*i.e.*, capable of exchanging only thermal energy with the surroundings).

Regardless of the nature of the energy from which it derives, the available work potential of a chemical substance at a particular combination of temperature and pressure is referred to either as the Gibbs energy or the Gibbs free energy. When expressed as an intensive property in terms of energy per mol, the Gibbs energy is referred to either as the Gibbs potential or the chemical potential. Differences in this potential between states of a substance and between substances in well-defined states are denoted by $\Delta G$. Since it is the difference in the intensive properties of a system that determines the driving force for change, the driving force of Reaction (1) as presented may be considered to be this difference between A and P as given by the van't Hoff reaction isotherm:

$$\Delta G = -RT\ln K + RT\ln(p/a) = \Delta G° + RT\ln(p/a) \qquad (2)$$

Here $K$ is the equilibrium constant, $\Delta G°$ is the so-called 'standard' difference in characteristic Gibbs free energy between one mol of A and one mol of P, and $\Delta G$ is the amount of Gibbs free energy consumed in the net conversion of one mol of A into one mol of P under a specific set of conditions of temperature, pressure, $a$, and $p$, the temperature, pressure, and other environmental conditions being the same as those used in the determination of $K$ and $\Delta G°$. Since $\Delta G$ as a driving force always refers to free energy that is consumed, it is usually presented as $-\Delta G$ (*i.e.*, as a positive quantity) when used in this sense.

The van't Hoff reaction isotherm serves to acknowledge that a particulate substance that is capable of undergoing chemical change possesses (*i*) an available work potential that is independent of its concentration and is characteristic of the substance in a given environment, and (*ii*) an available work potential that depends only on the number of individually mobile particles per unit volume (*i.e.*, only on particle concentration). The isotherm, as it applies to Reaction (1), can be derived from the ideal gas equation on the basis of the observation by van't Hoff [1] that the free energy of a solute in dilute (ideal) solution depends on the osmotic pressure of the solute just as the free energy of an ideal gas would depend on its pressure. Since, for an isothermal change in the state of such a gas, the pressure and concentration would vary in direct proportion to one another, the expression for a same-temperature difference in free energy of concentration in terms of pressure

*Nature of Free Energy and Entropy*

obtained from the ideal gas equation may be modified by substituting concentration for pressure. The expression then can be used as a measuring device to determine differences in characteristic chemical potential energy between the reactants and products of chemical reactions conducted in solution at constant temperature and pressure from equilibrium differences in translational kinetic energy density (*i.e.*, from equilibrium differences in free energy of concentration). The means by which this is ordinarily achieved in respect to biochemical thermodynamics, in which case the solutions are usually considered to be ideal, is outlined immediately below.

**Determination of $\Delta G^\bullet$ by the Equilibrium Method.** At a given temperature and pressure, the Gibbs potential $G$ of a chemical substance in ideal solution is equivalent to $RT\ln c + C$, where $c$ is the concentration and C is an unknown constant. As is evident from the fact that it can be expressed in the form $V = Ae^{-G/RT}$, where $V = \frac{1}{c}$ = molar volume and $A = e^{C/RT}$, this foundational relationship of the classical approach to chemical thermodynamics has much in common with those of the Maxwell-Boltzmann statistical approach. In consequence of C being unknown, $G$ must be determined in relation to a reference Gibbs potential $G°$ and its concentration component $RT\ln c°$, which, by convention, are subtracted from $G$ and $RT\ln c$. Thus, $G - G° = RT\ln c - RT\ln c°$ and $G = G° + RT\ln(c/c°)$. By convention, the difference in Gibbs potential between a reactant and a product, such as A and P of Reaction (1), is obtained by subtracting the Gibbs potential of the reactant from that of the product. Thus, for Reaction (1):

$$[G_P - G_A] = [G_P° - G_A°] + [RT\ln(p/p°) - RT\ln(a/a°)]$$

$$\Delta G = \Delta G° + RT\ln(p/a) - RT\ln(p°/a°) \quad (3)$$

As may be seen, acquisition of the reaction isotherm in the form (Equation 2) in which it is usually presented and used in respect to Reaction (1) requires that we get rid of the $RT\ln(p°/a°)$ term in Equation (3). Since the logarithm of unity is zero, this can be done very easily in this particular case by assuming $a° = p°$. By making this assumption, we eliminate the free energy of concentration (mechanical) component of $\Delta G°$ and thereby assume $\Delta G°$ to be equivalent to the difference only of the characteristic (chemical) kind of Gibbs free energy between one mol of A and one mol of P.

Since A and P are assumed to be in the same environment, the difference in characteristic free energy on a per mol basis between them in their common environment could be determined in a closed system from their relative concentrations when the reaction is at equilibrium, in which case $\Delta G$ would be zero, $p/a$ would be equivalent to the equilibrium constant of the reaction, and $\Delta G°$ would be equivalent to $-RT\ln K$. This assumes of course that the equilibrium constant is not so large or small as to preclude measurement of $a$ or $p$.

As may be seen by including the reference concentrations in the reaction isotherm, the equilibrium constant is invariably a dimensionless quantity. Such is required if the logarithm of an equilibrium constant is to make sense [46]. Inclusion of the reference concentrations in the isotherm has the additional advantage of allowing one to see clearly how to handle a reactant or product that is maintained at constant concentration or virtually so in the course of the reaction. Thus for such a reactant or product, one need only specify the constant concentration to be the reference concentration, in which case the reactant or product would be eliminated from the isotherm. Of course any special condition of this nature would limit the equilibrium constant to that condition and it would thus be necessary to specify the limitation in presenting the equilibrium constant if not done so by convention.

For reactions having multiple reactants and products, one's choice of reference concentrations is arbitrary within the restriction that the sum of the reference free energies of concentration on a per mol basis be the same for the reactants as for the products. Note that, since Gibbs energies are expressed on a per mol basis for each mol of each reactant and product specified in the stoichiometric equation for the reaction, reactions having more than or less than one mol of a particular reactant or product in their stoichiometric equation must be treated such that $\Delta G°$ for the reactions $2A \rightleftharpoons P$ and $\frac{1}{2}A \rightleftharpoons P$, for example, be equivalent to $G_P° - 2G_A°$ and $G_P° - \frac{1}{2}G_A°$, respectively. Note also that, regardless of the complexity of the reaction, $\Delta G°$ is expressed in units of energy per mol. For reactions other than the very simple Reaction (1), one might ask: energy per mol of what? By taking into account the ideal-gas and mechanical (Figure 1) origins of the above method and using reciprocal concentrations (*i.e.*, molar volumes) rather than concentrations, one can readily see that the answer must be: per mol of individually mobile particles. In other words, one can readily see that the characteristic Gibbs energy of each mol or fraction of a mol of each reactant and product specified in the stoichiometric equation is measured in terms of the density of the translational kinetic energy possessed by one mol of individually mobile particles at the temperature of the reaction system. Thus, the 'per mol' may be interpreted to indicate simply that the stoichiometric coefficients refer to mols rather than to molecules.

The conventional way of eliminating the reference term is to assume for each kind of reactant and product in the stoichiometric equation a hypothetical 'standard state' concentration of unity in whatever units of concentration are employed. This state is 'hypothetical' in that the reactants and products are assumed to have properties identical with those they would have if they were in dilute (ideal) solution, in which case the requirement that concentrations be in terms of numbers of individually mobile particles per unit volume is likely to be met. Although this convention invariably abolishes the reference concentration term and, since the logarithm of unity to any power is zero, is the most logical and efficient means of doing so, the fact of its use is often stated in such a way as to give the very confusing impression that the reactants and products must be at a concentration of one molar or one molal, for example, if the value of $\Delta G°$ is to be correct, implying incorrectly that its use has a purpose in addition to specification of dimensional units and elimination of the reference term.

**Nature of Characteristic Chemical Potential Energy.** As implied above, the characteristic component of the free energy of a chemical substance consists of all forms of available work potential other than the intrinsically mechanical free energy of concentration. From the standpoint of chemistry, the characteristic free energy of a substance at a particular temperature consists only of that associated with what is ordinarily referred to as chemical bond energy. Bond energies are usually dealt with in terms only of thermal energies of bond formation or dissociation. This being the case, it is necessary to inquire as to the general nature of bond energy as a nonthermal actual energy. Since, other than the extremely weak gravitational force, only Coulomb (electrostatic) forces and forces deriving therefrom are known to exist between atoms [47], bond energy as a nonthermal actual energy must consist of these forces and, since binding energy in respect to a system is defined as the net energy required to decompose the system into its constituent particles, must correspond to what is commonly referred to as chemical binding energy.

Since atoms and molecules possess both attractive and repulsive forces, and since the formation of a kinetically stable chemical bond at a finite temperature invariably involves the establishment of a balance between these forces within an energy barrier to the making and breaking of the bond, chemical binding energy must refer to the energies of both attractive and repulsive forces. Consider, for example, two kinds of charged atoms $A^+$ and $A^-$ which possess binding energy only by virtue of their net charges. Since unlike charges attract one another, binding energy for a binding interaction between $A^+$ and $A^-$ must necessarily be viewed as being attractive. On the other hand, since like charges repel, binding energy for an interaction between $A^+$ and $A^+$, for example, must necessarily be regarded as being repulsive. Since the attractive binding interaction would clearly involve a net consumption of binding energy, the repulsive one must by implication involve a net production of binding energy. Since both attractive and repulsive binding energy must be forms of nonthermal actual energy, it appears from this example that the binding energies of molecules possessing particularly large amounts of bond energy are likely to be of the repulsive kind. However, as is evident from the fact that the bulk of local matter occurs naturally as liquids and solids consisting of attractively bound molecules, most binding interactions between atoms are net attractive. In view of this and of the large amount of energy required to force like charges into close proximity to one another, it seems likely that most of the molecules we ordinarily think of as possessing relatively large amounts of bond energy are molecules in which the constituent atoms are relatively unattractive to one another and/or are constrained (bonded) in such a way as to prevent optimal neutralization of their attractive forces. If we were to consider this invariably to be the case, we would be ignoring the fact that, by virtue of the existence of energy barriers to the making and breaking of chemical bonds, it is possible for kinetically stable bonds to be formed between atoms despite the interaction being net repulsive.

As is evident from the fact that most chemical reactions that are thermodynamically favorable in respect to $\Delta G°$ are exothermic, translational kinetic (mechanical) energy of atoms and molecules can be constrained as bond energy. Uncatalyzed conversions of translational kinetic energy into bond energy must occur through collisions of the atoms or molecules with sufficient energy to force the electronic and other changes (*e.g.*, desolvations) that constitute the energy barrier. From the fact that bound atoms and molecules tend to dissociate as temperature is increased, it is evident that the collision energy can be too high as well as too low.

Since the amount of attractive binding energy possessed by a substance at constant ambient pressure depends on its temperature as well as on its chemical composition, it is necessary to distinguish between changes in attractive binding energy that do and do not constitute changes in characteristic chemical potential energy. This distinction was in effect made above by considering thermally produced attractive binding energy to be extramolecular latent thermal energy. According to the treatment given, the attractive binding energy that accumulates in an individual substance as its temperature is increased at constant pressure from absolute zero to the point at which all the molecular constituents are sufficiently free to be individually mobile is not characteristic chemical potential energy. That such is the case is evident from the above consideration of a reversible conversion of ice into liquid water with thermal energy deriving from nonthermal actual energy used to compress an ideal gas. As was noted, the free energy of the nonthermal actual energy would remain with the gas rather than undergo transfer with the thermal energy that would become latent thermal energy of newly formed liquid.

One can reasonably expect single substances to possess latent thermal energy consisting of attractive binding energy having the quality of characteristic chemical potential energy only if the substances are in supercooled states. Thus, whereas a substance in a superheated state would be out of thermochemical equilibrium as a result of possessing an excess of translational kinetic energy relative to attractive binding energy, a substance in a supercooled state would be out of thermochemical equilibrium as a result of possessing an excess of attractive binding energy relative to the mechanical energy. That the excess attractive binding energies of substances in supercooled states actually have the quality of characteristic chemical potential energy is particularly evident from studies on single solids that have been mechanically or otherwise reduced to nano-sized particles at temperatures far below their melting points. Single solids thus modified are in effect in extremely supercooled states and have chemical reactivities and other properties, noted above, indicative of the presence of exceptionally large amounts of characteristic chemical potential energy [48, 49]. For instance, graphite reduced to extremely small particles by prolonged grinding under inert-gas conditions has been observed to be sufficiently reactive to ignite spontaneously at room temperature when exposed to air [50].

**Determination of $\Delta G^\bullet$ by the Calorimetric Method.** Since consumptions and productions of bond free energy under the constant temperature, constant pressure, and virtually constant volume conditions assumed here are accompanied by productions and consumptions, respectively, of equivalent amounts of energy having the quality of thermal energy at the specified constant temperature, differences in characteristic Gibbs free energy between the reactants and products of chemical reactions can be determined in closed systems not only by the equilibrium method, but also by measuring thermal energy changes occurring as a result of the reactions. However, this 'isothermal' calorimetric method is much more difficult and less likely to be feasible in that it requires a means of distinguishing between thermal energy changes due to bond energy changes and irrelevant ones due to interconversions between the latent and nonlatent thermal energies. An irrelevant change will occur with any net change in the heat capacity of the reaction system. Such a change could occur as a result primarily either of a change in the total intramolecular latent thermal energy of the system or of a change in the total free energy of concentration (*i.e.*, in the total number of individually mobile particles in the system). As may be seen by comparing the reactions $A \rightleftharpoons P + Q$ and $A_2 \rightleftharpoons 2A$, a net increase in the total free energy of concentration due to a chemical change under the conditions assumed here will be accompanied by an increase in a latent thermal energy that will be indistinguishable from if not identical with the above-described extramolecular kind.

That free energy of concentration should be viewed here as elsewhere as being intrinsically mechanical rather than chemical can be seen by considering a closed Reaction (1) system in which A and P possess equal amounts of bond energy on a per mol basis and the reaction does not involve a net change in the system as to latent thermal energy. In such a system, an interconversion between A and P would not involve a net change in an actual (First-Law) energy of any kind of either the system or its surroundings. Nevertheless, an irreversible net conversion would occur and thus be driven with consumption of free energy if there were a differential in the concentrations of A and P. Since the system is assumed to be closed, a differential between A and P as to concentration would be accompanied by differentials as to absolute amount of both bond energy and translational kinetic energy. Since, for a given differential in concentration, the magnitude of the driving force for the reaction would be a function only of temperature, it is evident that, of these two kinds of actual energy, only the thermal energy would be relevant in respect to the driving force. However, the fact of the reaction not being accompanied by a change in actual energy would rule out the thermal energy as well as any other actual energy of the system as the source of the free energy consumed in driving the reaction. As will be pointed out below by comparing Reaction (1) systems with the purely mechanical Figure 1 system, the source of the free energy would be the nonthermal actual energy used to generate the differential in concentration. Since only free energy from this source could be present in the closed system, it would be necessary to attribute the driving force for the reaction simply to the differential in concentration or 'number density' between A and P. Accordingly, the reaction would be driven as a result of there being a differential between the number of A molecules converting spontaneously into P molecules and the number of P molecules converting spontaneously into A molecules.

Consider now a $\Delta G° = 0$ reaction which differs from the one above in that it involves a net change in latent thermal energy and thus a net change in heat capacity. Since the reaction would not involve a net change in bond energy and the change in heat capacity and latent thermal energy would occur only in the system, maintenance of the system at constant temperature would require transmission of thermal energy between the system and its surroundings. Since for a closed system such transmission could be driven only by a gradient of temperature, and since by definition any latent thermal energy produced or consumed at a particular temperature would have the same quality as nonlatent thermal energy at that temperature, an interconversion between these energies in such a system maintained at constant temperature would clearly be irrelevant to the energetics of the chemical changes that constitute the reaction. In other words, as a result of the latent and nonlatent thermal energies having different locations and the same quality at any particular temperature, an interconversion between these energies in such a system could be driven only by a gradient of temperature, which, in consequence of the constant-temperature specification, would by definition be sufficiently small as not to result in a finite consumption of free energy. From this it is evident that, from the standpoint of theory, a change in latent thermal energy in any particular reaction system of the sort considered here could be determined directly from a detectable change in thermal energy only if the reaction could be conducted reversibly. Consequently, for a reaction involving a significant change in latent thermal energy, the only way to obtain a reasonably accurate estimate of $\Delta G°$ by the calorimetric method would be to determine the difference between the amount of nonlatent (detectable) thermal energy produced or consumed in the course of the reaction when the reaction is conducted 'completely irreversibly' (*i.e.*, under conditions of zero work being done on the surroundings) and the amount produced or consumed when the reaction is conducted very nearly reversibly (*i.e.*, under conditions in which very nearly all of the work potential that becomes available in the course of the reaction is used to do work on the surroundings) (*cf.* [51, 52]).

For a reaction system maintained at constant temperature and volume rather than at constant temperature and pressure, the detectable change in thermal energy occurring under conditions of complete irreversibility would correspond to the algebraic sum of the changes in the chemical and latent thermal energies of the system. A change of this sort on a per mol basis is usually denoted by $\Delta E°$ and referred to as the difference or change in the total energy of the system. The detectable change in thermal energy occurring under conditions of near reversibility would correspond closely to the change in the latent thermal energy of the system in the opposite direction. Changes of this sort on a per mol basis are denoted by $T\Delta S°$. The change in bond energy on a per mol basis would be given in terms of thermal energy by the difference $\Delta E° - T\Delta S°$ and would correspond closely to the difference between the reactants and products in respect to characteristic Helmholtz free energy.

Changes in characteristic Gibbs free energy are reckoned in the same way except that $\Delta E°$ is replaced by $\Delta E° + P\Delta V°$ and referred to as the difference or change in enthalpy, denoted by $\Delta H°$. The $P\Delta V°$ component of $\Delta H°$ serves to account for the pressure-volume work that would be done on the surroundings (atmosphere) by the system or by the surroundings on the system if the reaction were to result in an increase or decrease, respectively, in the volume of the system. In consequence of the $P\Delta V°$ component, $\Delta G°$ for a Reaction (1), for example, could correspond exactly to the difference in bond energy between one mol of A and one mol of P in their common environment only if $\Delta V°$ for the conversion should happen to be zero. Although changes in volume ordinarily occur, they are usually small for reactions involving net changes only in small molecules in solution at constant temperature and pressure. At normal (atmospheric) pressures the amounts of energy involved are usually also small and sufficiently so to justify considering them to be negligible in relation to the change in bond energy.

The above means of estimating and accounting for differences in the characteristic component of the Gibbs free energy is in effect a means of doing so in terms of changes in thermal energy of both the system and its surroundings in terms only of system properties. Thus:

$$\Delta G°_{sys} = \Delta H°_{sys} - T\Delta S°_{sys} \qquad (4)$$

$$-\Delta G°_{sys}/T = -\Delta H°_{sys}/T + \Delta S°_{sys} = \Delta S°_{surr} + \Delta S°_{sys}$$

$$-\Delta G°_{sys} = T(\Delta S°_{surr} + \Delta S°_{sys}) = T\Delta S°_{total}$$

**Enthalpy-Entropy Compensation.** The $\Delta H°$ and $T\Delta S°$ components of the above 'Gibbs standard free energy relation' (Equation 4) are often referred to as differences or changes in energy and entropy, respectively, and are often referred to in such a way as to give the impression that the 'energy' and 'entropy' are different forms of energy that are more or less equally capable of driving thermodynamic processes under conditions of constant temperature and pressure. That doing this is not appropriate is evident from the fact that, since $\Delta H°$ is equivalent to the algebraic sum of $\Delta G°$ and $T\Delta S°$ and serves to represent the difference or change in the total energy of the system, $T\Delta S°$ must represent a difference or change in an energy of the system that is independent of the difference or change in thermal energy represented by $\Delta G°$ and has the quality of thermal energy at the temperature of the system. In other words, $T\Delta S°$ must serve as indicated above to represent a difference or change in the kinds of energy referred to as latent thermal energy. Accordingly, the result $\Delta G° = -T\Delta S°$ for an isothermal process would mean simply that the change in thermal energy due to the change in the characteristic Gibbs free energy of the system happens to be accompanied by an equivalent but oppositely directed change in the latent thermal energy of the system, and the result $\Delta G° = \Delta H°$ would mean simply that the process does not involve a net change in the latent thermal energy of the system.

Several problems with current and past interpretations of the Gibbs standard free energy relation have been pointed out and/or discussed in recent years by a number of investigators (see, *e.g.*, [53-60]). Some of the problems derive from the fact that reactions occurring in solution are ordinarily accompanied by a 'solvent reaction' which is not accounted for in stoichiometric equations and which in some cases appears to make equivalent contributions of the same sign to $\Delta H°$ and $T\Delta S°$ and thus appears to make little or no contribution to $\Delta G°$ [53, 61-66]. For reactions occurring in aqueous solution, the energy of this so-called 'enthalpy-entropy compensation' can be quite large as to amount and, in such cases where the associated change in $\Delta G°$ is very small, obviously must derive from large conversions between the latent and nonlatent thermal energies somewhat as in the case of reversible conversions between the solid and liquid forms of water. In this regard, it is important to note that, since $\Delta H°$ serves to represent the change in total energy and thus includes $T\Delta S°$, enthalpy-entropy compensation in a broad sense must invariably occur in processes for which $T\Delta S°$ is finite.

Although widely acknowledged to be a phenomenon that is not well understood, enthalpy-entropy compensation of the solvent-dependent kind is generally thought to occur as a result of differences between reactants and products with respect to strength of binding interaction with the solvent and thus with respect to amount of solvent bound or otherwise immobilized. Such is consistent with the fact that binding interactions between solutes and solvents can reasonably be expected to follow the same rules as binding interactions between solutes. Thus, binding interactions between molecules of a solute and its solvent to form complexes can be either favorable, with net conversion of binding energy into thermal energy, or unfavorable, with net conversion of thermal energy into binding energy. Also, since particles of an undissolved solute ideally immersed in a solvent would be completely surrounded by the solvent, one could expect there to be a considerable mechanical (mass-action) driving force ($-\Delta G_{conc}$) for solvation of the solute from the high solvent activity alone. Thus, one could expect some solvation of the solute to occur even if the attractive forces between the solute and solvent molecules should be very weak. If the attractive forces between molecules of the solute and solvent should be weaker than those between molecules of the solvent, one could expect solvation of the solute to be accompanied by net conversion of thermal energy into attractive binding energy due to hindrance of bond formation between solvent molecules more or less as proposed by Hildebrand and coworkers [67, 68], resulting in a temperature-sensitive tension in the solvent at the surfaces of the solute molecules similar to that which occurs at the interface of the solvent and its vapor. As a result of this tension, one could expect the existence of a force for minimization of the amount of space occupied by the solute. Accordingly, surface tension in a liquid results from molecules at the surface possessing relatively large amounts of attractive binding energy which in turn results in there being a relatively large attractive (contractive) force among molecules at the surface that tends to minimize surface area [69].

Consistent with these expectations, solvations of nonpolar (hydrophobic) solutes in water tend to be thermodynamically unfavorable with respect to $\Delta G°$ and the solubilities of such solutes in water tend to be lower as the surface areas of the molecules are larger (for recent reviews, see [70-72]). In addition, water has an exceptionally high surface tension which decreases with increase of temperature [69], and solvations of nonpolar solutes in water are generally accompanied by a temperature-sensitive decrease in the amount of space occupied by the solute [73-75]. In contrast to $\Delta G°$, both $\Delta H°$ and $T\Delta S°$ for solvation tend to be negative, indicating that the latent thermal energy of the system tends to decrease more than the bond energy increases. The large negative $T\Delta S°$ associated with the above phenomena, widely referred to collectively as 'the hydrophobic effect', is usually attributed to a temperature-sensitive ordering of water molecules at the surfaces of the solute molecules [75-78]. Accordingly, solvation is ordinarily accompanied by an increase in heat capacity that can be readily explained in terms of reversal of the solvent ordering as the temperature is increased for the purpose of measuring the heat capacity [70]. Consistent with the above, 'Hildebrand' interpretation of the hydrophobic effect, the ordering appears to be limited to a single layer of water molecules [79].

Contrary to the Hildebrand view, the postulated ordering of water and the observed increase in bond energy in the hydrophobic effect have generally been interpreted to mean that there is an increase in hydrogen bonding, either in number of bonds or in bond strength, and that the thermodynamics of the ordering process are thus anomalous [72]. If we were to consider the ordering actually to involve an increase in the number of hydrogen bonds, we could expect it to be accompanied by a decrease in bond energy rather than by the observed increase and the thermodynamics would in fact appear to be anomalous, since hydrogen bonding occurs through attractive (cohesive) forces and therefore can be expected to consume binding free energy. On the other hand, if we were to consider the ordering to involve only an increase in hydrogen bond strength and were to interpret 'increase in bond strength' to mean 'increase in bond energy', we could conceive of the thermodynamics of the ordering process not being anomalous and of there being no actual disagreement between the conventional and Hildebrand views. Thus it is conceivable that the ordering involves only a net increase in the amount of bending and stretching of hydrogen bonds, in which case there could be an increase in attractive binding energy without a net decrease in the number of bonds. This possibility is based on the fact that, by virtue of the existence of energy barriers to the making and breaking of chemical bonds, the energy of a chemical bond must be enhanced if the bond is to be broken. In the case of sustained bending or stretching without net bond breakage, 'enhancement energy' would be retained somewhat as in the case of the bonds in a stretched rubber band and would consist exclusively of attractive binding energy. As will be outlined below by considering a solvent reaction in respect to Reaction (1), the large negative $T\Delta S°$ and the associated solvent-dependent enthalpy-entropy compensation in the hydrophobic effect can be explained readily and operationally in terms of the postulated ordering of solvent molecules simply by taking into account the fact that the ordering implies immobilization.

In doing this, it shall be assumed (*i*) that water is the solvent, (*ii*) that molecules of the solvent bind less strongly to each other than to A and P, and (*iii*) that the attractive binding interaction between the solvent and P is relatively strong. This being the case, we could expect a net conversion of A into P to be accompanied by a solvent reaction involving a net conversion of mobile water into an immobile, bound form and an associated net conversion of attractive binding energy into thermal energy. In a closed Reaction (1) system, we could expect this conversion of binding energy into thermal energy to make a negative contribution to the $\Delta G°$ for the reaction just as would any other net conversion of binding energy into thermal energy at constant temperature and pressure. Therefore, we could expect the contributions to $\Delta H°$ and $T\Delta S°$ to differ accordingly. How, then, can a difference between the reactant and product with regard to solvent immobilization account for solvent-dependent enthalpy-entropy compensation? As surmised by Ives and Marsden [62], this question can be answered by taking into account changes in a specific extensive (size) variable of closed reaction systems.

Enthalpy-entropy compensation of the above sort can be adequately explained by invoking the above-noted prediction of the equipartition principle that, at any particular temperature, the individually mobile particles of a gas will possess on average the same amount of translational kinetic energy regardless of the nature of the particles. That this prediction is generally considered to be correct and applicable to substances in ideal solution is evident from the fact that it is implicit in the concentration term of the reaction isotherm (Equation 2) through the assumption that, at any particular temperature, only the relative concentrations of A and P as particles matters in regard to the difference in free energy of concentration. According to this assumption, differences in the characteristic features of A and P, such as size (mass), net charge, amount of bond energy, amount of bound solvent, and amount of intramolecular latent thermal energy, are of no consequence whatever as regards the contributions of A and P to free energy of concentration. In regard to particle size, it may be noted that Einstein [80] employed the above assumption in his very successful treatment of Brownian movements of particles of sufficient size to be seen by light microscopy. This and the fact that one could reasonably expect to be capable of demonstrating the equilibrium value of $RT\ln(p/a)$ for a Reaction (1) as a difference in mechanical (osmotic) work potential equivalent to the $\Delta G°$ for the reaction imply (*i*) that differences in free energy of concentration are differences in the density of translational kinetic energy and (*ii*) that, at any particular temperature, individually mobile particles of A and P would possess on average over time equivalent amounts of translational kinetic energy regardless of their characteristic features. This being the case, one could reasonably expect equipartition of the translational kinetic energy of a closed reaction system among all the particles of the system, including, of course, those of the solvent.

Accordingly, if P of a closed Reaction (1) system were to immobilize more water than does A, one could expect a net conversion of A into P to be accompanied by (*i*) a decrease in the number of particles in the system, (*ii*) an equipartitioning of the total translational kinetic energy of the system among the smaller number of particles, (*iii*) an augmentation of the average translational kinetic energy of the particles, (*iv*) an elevation of the temperature of the system, and (*v*) a transmission of thermal energy from the system to the surroundings. Since the decrease in number of particles would occur regardless of the driving force for the conversion of A into P and thus regardless of whether the reaction is conducted reversibly or irreversibly, it would make a negative contribution to both $\Delta H°$ and $T\Delta S°$. Of course if A were to immobilize more water than does P, the sign of the contribution would be positive rather than negative. In either case, the contribution could be regarded as being due to a change in free energy of concentration that is not accounted for in the stoichiometric equation for the reaction.

Since the binding of water to solute and of water to water can be expected to vary with temperature, one could expect the magnitude of the contribution to be temperature dependent. For reactions involving net changes in large biological molecules, in which cases changes in latent thermal energy on a per mol basis tend to be particularly large, this dependence is particularly noticeable. A question currently of interest in this regard is why applications of the integrated form of the van't Hoff equation:

$$\ln K = -\Delta H°/RT + \text{constant} \qquad (5)$$

to reactions of the above sort do not yield correct values for $\Delta H°$ despite yielding linear plots of $\ln K$ vs. $1/T$ [54-60]. This apparent discrepancy can be resolved by noting that, since $\ln K$ is also equivalent to $-\Delta G°/RT$, the integration constant of Equation (5) must be equivalent to $\Delta S°/R$ if the equation is to be consistent with Equation (4), and that, since the validity of Equation (4) is beyond dispute, the right-hand side of Equation (5) must therefore be equivalent to $-\Delta G°/RT$. This being the case, linearity of plots of $\ln K$ vs. $1/T$ means nothing more than that $\ln K$ is directly proportional to $-\Delta G°/T$, the observed proportionality constant being an estimate of $1/R$, and that $\Delta G°$ is a linear function of $T$.

**Role of Free Energy of Concentration in Free Energy Transfer and Conservation.** As was noted above, the nature of free energy of concentration is fundamentally the same for a system in which only chemical work is possible as for an ideal gas system in which only mechanical work is possible. Accordingly, the concentration term of the reaction isotherm ignores completely the presence of the solvent. Thus, $RT\ln(p/a)$ of Equation (2) would be applicable to a closed Reaction (1) system even if A and P were gases in an ideal gas-phase reaction. This being the case, we can apply to a closed Reaction (1) system virtually everything said above concerning the energy and entropy changes of an ideal gas associated with compression and expansion of the gas between two states at the same temperature simply by ignoring the means by which the two states were assumed to be achieved and considering $a$ and $p$ to be the two states. The fact that the volume would be essentially constant for the two states in the closed Reaction (1) case would not matter because the difference in free energy of concentration between them on a per mol basis could be determined from knowledge of the relative values of $a$ and $p$. The fact that A and P would share the same space also would not matter because A and P would be virtually independent of one another in the ideal case and thus may be thought of as being individual systems within an overall Reaction (1) system.

The individual systems of a closed Reaction (1) system may be thought of as being pitted against one another in respect to compression and expansion. For instance, if $a$ and $p$ were equal and A possessed $RT\ln 10$ more bond energy than P on a per mol basis, the A system, through 'expansion', would be capable of 'compressing' the P system to the extent that $p$ exceeds $a$ by a factor of ten, at which point (*i*) the reaction would be at equilibrium, (*ii*) the amount of bond energy in the individual systems would be the same on an absolute amount basis, and (*iii*) the A and P systems would differ as to free energy of concentration on a per mol basis by $RT\ln 10$.

Since at equilibrium the absolute amount of bond energy would be the same for the A and P systems, the difference in free energy of concentration, although generated as a result of there having been a difference in bond energy on an absolute amount basis, obviously would not be associated with such a difference at equilibrium. Therefore, since any consumption of bond free energy in the course of the reaction going to equilibrium would be accompanied by the production of an equivalent amount of thermal energy, and since all of this thermal energy could depart the system as a result of the temperature being maintained constant if the conversion of A into P should happen not to result in an increase in the latent thermal energy of the system, the free energy of concentration generated in the P system, like that which one could expect to be generated in an ideal gas upon its isothermal compression, could appear to be free of the actual energy that undergoes transformation in the course of its generation.

We turn now to the question of how the difference in free energy of concentration could be utilized to do chemical work (*i.e.*, to convert ambient thermal energy into chemical bond energy). As noted above, Reaction (1) as presented is assumed to be linked to others through A being a product of a preceding reaction and through P being a reactant of a succeeding one in a stationary-state, biological-type system. To rule out the possibility of differences in free energy of concentration being used to do mechanical (osmotic) work in addition to chemical work, the individual reactions of this system shall be assumed to be located in the same com-partment. If Reaction (1) were thermodynamically favorable in respect to $\Delta G°$, a difference in free energy of concentration of the above sort could be generated and conserved as bond free energy through the differential between *a* and *p* either pulling preceding reactions or pushing succeeding ones that are thermodynamically unfavorable in respect to $\Delta G°$ and thus require thermal energy to proceed in the same direction as Reaction (1). If the individual reactions were coupled only by freely diffusible intermediates, transmissions of actual energy between the reactions would occur only through transmissions of thermal energy. Of course the role of the differential in free energy of concentration in conserving the excess bond free energy of A relative to P on a per mol basis would be to increase the number of reactant molecules converting spontaneously into product molecules relative to the number of product molecules converting spontaneously into reactant molecules in the unfavorable reactions. If the stationary-state ratio of *p* to *a* were infinitesimally smaller than the equilibrium constant of Reaction (1), all of the reactions would be operating reversibly in the thermodynamic sense and all of the bond free energy would be conserved as bond energy.

Although the biological system as one in which only chemical work is possible is similar to the purely mechanical (Figure 1) system in that it employs a difference in density of translational kinetic energy to mediate the transfer of free energy from one repository to another, it differs markedly from the purely mechanical system in that it employs the particle-number-density (concentration) aspect rather than the mechanical-energy-density aspect of the translational energy density. Owing to the need for thermal energy transfer between the system and surroundings at a low temperature differential to conserve free energy in the purely mechanical case and the lack of such need to conserve free energy in the biological case, the two systems also differ greatly as to the amount of time that would be required for efficient transfer of the actual energy. Thus, in the biological case the reactions yielding and requiring thermal energy could occur simultaneously in close proximity to one another at the microscopic level, a consequence of which would be that the requisite thermal energy transfers could occur rapidly while the temperature of the system is essentially constant. Thermal energy transfer between the biological system and its surroundings would be needed only to rid the system of the thermal energy produced as a result of the need for expenditure of bond free energy to maintain an appropriate net rate of reaction. If the system should require thermal energy to maintain its temperature, this expenditure of free energy would also serve to meet this need.

**APPENDIX**

According to the main body of the text, by considering irreversibility in respect to an adiabatic cycle of compression and expansion of an ideal gas, one can readily see that any irreversibility in a thermodynamic work process would be reflected only in the characteristic entropy and would result in the consumption of an amount of free energy equivalent to an amount of ambient thermal energy produced. That these observations are correct may be seen from the following considerations in reference to an adiabatic version of the Figure 1 model of the main body of the text.

If one mol of an ideal gas in the fully expanded state $V_1$ at a temperature $T_1$ of the reservoir were compressed to a volume $V_2$ adiabatically, the temperature of the gas would increase by an amount depending on the heat capacity and on the amount of energy $\Delta E$ imparted to the gas. This amount of energy would be equivalent to the amount of work $W$ done and thus would depend on the rate of the compression. Regardless of the rate, $W$ and $\Delta E$ would be equivalent to $C_V \Delta T$. Since $S$, $T$, and $V$ are state functions and $C_V$ would be a constant, the expression for the net change in entropy

$$\Delta S_{net} = \Delta S_{char} + \Delta S_{conc} = C_V \ln(T_2/T_1) + R \ln(V_2/V_1)$$

in terms of these parameters would be applicable regardless of the rate of the compression. If the gas were compressed reversibly, $\Delta S_{char}$ would be equivalent to $-\Delta S_{conc}$ and $T_2$ could be determined from the relationship $T_2 = T_1(V_1/V_2)^{R/C_V}$. If the gas were compressed at a finite rate, $T_2$ would be relatively high and $\Delta S_{net}$ for the compression would be finite and positive as a result of the increase in characteristic entropy exceeding the decrease in entropy of concentration. In this 'irreversible' case, $T_2$ could be determined only by experimental means. Although calculation of $\Delta S_{net}$ could be easily achieved if $T_2$ for the irreversible compression were known, doing so would not be helpful in respect to precise quantification of the amount of free energy consumed. However, since the work potential of the gas would increase by an amount equivalent to the amount of work done regardless of the rate of the compression, doing so could serve to indicate in a semiquantitative fashion that work potential has been rendered unavailable. As indicated below, if $T_2$ were known, the precise amount rendered unavailable could be determined theoretically by considering the gas to undergo reversible expansion to its original volume.

In contrast to what would be the case for an ideal gas compressed irreversibly between two states at the same temperature, reversible expansion of the adiabatically compressed gas would result in a fraction of the thermal energy produced as a result of the gas being compressed at a finite rate being recovered as work, the fraction being that indicated by the efficiency of a Carnot engine operating reversibly between $T_2$ and the lower temperature that would exist upon return of the gas to its original volume $V_1$. Since $\Delta S_{char}$ for the expansion would be equivalent to $-\Delta S_{conc}$, this lower temperature would be equivalent to $T_2(V_2/V_1)^{R/C_V}$ and would of course be higher than the temperature $T_1$ of the reservoir by an amount depending on how fast the gas was compressed. In addition to the difference in temperature, there would remain differences in the pressure, thermal energy, and characteristic entropy, the difference in characteristic entropy being equivalent to that existing at the end of the compression phase. Elimination of these differences would require removal of the adiabatic constraint so that an amount of thermal energy $C_V \Delta T$ could undergo transmission to the reservoir, $C_V \Delta T$ in this case being equivalent to the amount of free energy consumed as a result of the gas being compressed at a finite rate. Since the amount of the gas was specified to be one mol, this amount would be equivalent to $C_V [T_2(V_2/V_1)^{R/C_V} - T_1]$. The value thus obtained divided by the temperature of the reservoir would be the net increase in entropy of the reservoir and for the overall process. Since the increase in entropy for the overall process would differ from that incurred by the gas in the compression phase and could be determined quantitatively only through determination of the amount of thermal energy transmitted to the surroundings, quantification of the net change in entropy for the cyclic process would in effect require prior quantification of the net change in free energy. Since the transmitted thermal energy is assumed to retain its form and to distribute in a volume sufficiently large as not to contribute to temperature, the net changes in entropy and free energy for the overall process would be changes only in the concentration kinds.

**ACKNOWLEDGEMENT:** This work was supported in part by the Stoner Research Fund of the Ohio State University Development Fund.